\documentclass[10pt]{article}
\usepackage{graphicx}
\usepackage{amsmath}
\usepackage{amssymb}
\usepackage{caption2}
\begin{document}
\begin{center}
{\large {\bf \sc{  Analysis of the   $Z_c(4200)$ as axial-vector molecule-like state
  }}} \\[2mm]
Zhi-Gang  Wang \footnote{E-mail: zgwang@aliyun.com.  }   \\
 Department of Physics, North China Electric Power University, Baoding 071003, P. R. China
\end{center}

\begin{abstract}
In this article, we assume the $Z_c(4200)$ as the color octet-octet type axial-vector molecule-like state, and construct the  color octet-octet type axial-vector
current to study its mass and width with the QCD sum rules.   The numerical values $M_{Z_c(4200)}=4.19 \pm 0.08\,\rm{GeV}$ and $\Gamma_{Z_c(4200)}\approx 334\,\rm{MeV}$ are consistent with the experimental data $M_{Z_c(4200)} = 4196^{+31}_{-29}{}^{+17}_{-13} \,\rm{MeV}$
and  $\Gamma_{Z_c(4200)} = 370^{+70}_{-70}{}^{+70}_{-132}\,\rm{MeV}$, and support assigning  the $Z_c(4200)$  to be  the color octet-octet  type  molecule-like  state with $J^{PC}=1^{+-}$. Furthermore, we discuss the possible assignments of the $Z_c(3900)$, $Z_c(4200)$ and $Z(4430)$ as the diquark-antidiquark type tetraquark states with $J^{PC}=1^{+-}$.
\end{abstract}

 PACS number: 12.39.Mk, 12.38.Lg

Key words: Molecule-like   state, QCD sum rules

\section{Introduction}

In 2014, the Belle collaboration analyzed the $\bar{B}^0\to K^- \pi^+ J/\psi$ decays with
  the full $\Upsilon(4S)$ data sample
corresponds to $711\rm{fb}^{-1}$ data sample collected by the Belle detector at the asymmetric-energy $e^+ e^-$ collider, and observed a resonance (named $Z_c(4200)$) in the $J/\psi \pi^+$ mass spectrum  with a statistical significance
of more than $6.2\,\sigma$, the measured    mass and width are
$M_{Z_c(4200)} = 4196^{+31}_{-29}{}^{+17}_{-13} \,\rm{MeV}$
and  $\Gamma_{Z_c(4200)} = 370^{+70}_{-70}{}^{+70}_{-132}\,\rm{MeV}$, respectively \cite{Zc4200exp,YuanCZ4200}.
 The preferred assignment of the quantum numbers is $J^P = 1^+$.

In 2007, the  Belle collaboration   observed a distinct peak   in the $\pi^{\pm} \psi^\prime$ mass spectrum   in the $B^0\to K^{\mp} \pi^{\pm} \psi^{\prime}$ decays
with the statistical significance of  $6.5\,\sigma$ \cite{Belle-2007}.
 In 2014, the LHCb collaboration analyzed the $B^0\to\psi'\pi^-K^+$ decays    by performing a four-dimensional fit of the decay amplitude, and  provided the first independent confirmation of the $Z(4430)^-$
and established its spin-parity to be $J^P=1^+$ \cite{LHCb-1404}.

In 2013, the BESIII collaboration studied  the process  $e^+e^- \to \pi^+\pi^-J/\psi$ and observed a structure $Z_c(3900)$ in the $\pi^\pm J/\psi$ mass spectrum \cite{BES3900}. Then the Belle and CLEO collaborations  confirmed the existence of the $Z_c(3900)$ \cite{Belle3900,CLEO3900}. Although the quantum numbers
are not measured, the assignment $J^P = 1^+$ is favored  if the decay $Z_c(3900)^\pm \to J\psi \pi^\pm$ takes  place in relative $S$-wave.
In 2014, the BESIII collaboration studied  the process $e^+e^- \to \pi D \bar{D}^*$ and observed a distinct charged structure $Z_c(3885)$  in the $(D \bar{D}^*)^{\pm}$
 mass spectrum, the  assignment   $J^P=1^+$ is favored \cite{BES-3885}. We tentatively  identify the $Z_c(3900)$
 and $Z_c(3885)$ as the same particle according to  the uncertainties of the masses and widths. For more literatures on the $X$, $Y$, $Z$ mesons, one can consult the recent review \cite{XYZ-review}.

The quark constituents of the $Z_c(3900)$, $Z_c(4200)$ and  $Z(4430)$ are $c\bar{c}u\bar{d}$ or $c\bar{c}d\bar{u}$, there are three analogous decays,
\begin{eqnarray}
Z_c(3900)^\pm&\to&J/\psi\pi^\pm\, , \nonumber \\
Z_c(4200)^\pm&\to&J/\psi\pi^\pm\, , \nonumber \\
Z(4430)^\pm&\to&\psi^\prime\pi^\pm\, ,
\end{eqnarray}
which take place through fall-apart mechanism.
The mass differences are $M_{Z(4430)}-M_{Z_c(3900)}=576\,\rm{MeV}$ and $M_{\psi^\prime}-M_{J/\psi}=589\,\rm{MeV}$, so it is natural to assign
 the  $Z(4430)$ to be the first radial
excitation of the $Z_c(3900)$ \cite{Z4430-1405,Nielsen-1401,Wang4430}.
Naively, we expect the tetraquark states have large decay widths, the $Z_c(4200)$ and $Z(4430)$ are good candidates of the tetraquark  states according to the widths $\Gamma_{Z_c(4200)} = 370^{+70}_{-70}{}^{+70}_{-132}\,\rm{MeV}$ \cite{Zc4200exp} and
$\Gamma_{Z(4430)}=\left(172\pm13\,{_{-34}^{+37}}\right)\,\rm {MeV}$ \cite{LHCb-1404}. In Ref.\cite{Wang4430-1GeV}, we study the axial-vector hidden charm (and hidden bottom) tetraquark states in details with the QCD sum rules  and obtain the value $M_{Z(4430)}=(4.44 \pm 0.19)\,\rm{GeV}$.
In Ref.\cite{ChenZhu416}, Chen and Zhu study the vector and axial-vector charmonium-like tetraquark states with the QCD sum rules and obtain the value $M_{J^P=1^+}=(4.16 \pm 0.10)\,\rm{GeV}$. In Ref.\cite{ChenZhu-1501}, Chen et al assume the $Z_c(4200)$ to be  the axial-vector tetraquark state and calculate its decay width  with the QCD sum rules. In Ref.\cite{Wang4430-1GeV}, the QCD spectral densities are calculated at the special energy scale $\mu=1\,\rm{GeV}$, while  in Ref.\cite{ChenZhu416},   the QCD spectral densities are calculated by taking the $\overline{MS}$ masses $m_c(m_c)$ and choosing the vacuum condensates at the energy scale $\mu=1\,\rm{GeV}$. If the $Z_c(3900)$, $Z_c(4200)$ and  $Z(4430)$ have the same quantum numbers $J^{PC}=1^{+-}$, they cannot all be the ground state axial-vector   tetraquark state with the same quark structure.

The $X$, $Y$ and $Z$ mesons have been studied extensively by the QCD sum rules \cite{Wang4430-1GeV,ChenZhu416,ChenZhu-1501,No-formular}, but the energy scale dependence of the QCD spectral densities are not studied.   In previous works \cite{Wang4430,WangHuang3900,Wang-tetraquark,WangHuang-molecule,Wang-molecule}, we  explore the energy scale dependence of the QCD sum rules for the hidden charmed and hidden bottom tetraquark states and molecular states
in details for the first time, and suggest a  formula,
\begin{eqnarray}
\mu&=&\sqrt{M^2_{X/Y/Z}-(2{\mathbb{M}}_Q)^2} \, ,
 \end{eqnarray}
 with the effective masses ${\mathbb{M}}_Q$  to determine the energy scales of the  QCD spectral densities.

The quarks have color $SU(3)$ symmetry,  we can construct the tetraquark states according to the routine  ${\rm quark}\to {\rm diqiurk}\to {\rm tetraquark}$,
\begin{eqnarray}
(3\otimes 3)\otimes(\overline{3}\otimes \overline{3})&=&(\overline{3}\oplus 6)\otimes(3\oplus\overline{6}) =\overline{3}\otimes3\oplus\cdots =1\oplus 8\oplus \cdots \, ,
\end{eqnarray}
or construct the molecular states and molecule-like states according to the routine ${\rm quark}\to {\rm meson\,\, (or\,\, meson-like\,\,\,state)}\to {\rm molecular\,\, state\,\, (or\,\, molecule-like\,\, state)}$,
\begin{eqnarray}
(3\otimes \overline{3})\otimes(3\otimes \overline{3})&=&(1\oplus 8)\otimes(1\oplus8) =(1\otimes1)\oplus (8\otimes8)\oplus\cdots =1\oplus 1\oplus \cdots \, ,
\end{eqnarray}
where the $1$, $3$ ($\overline{3}$), $6$ ($\overline{6}$) and $8$ denote the color singlet, triplet (antitriplet), sextet (antisextet) and octet, respectively.

In the  scenario of tetraquark  states, we study the $\overline{3}\otimes3$ type  (or the diquark-antidiquark type) scalar, vector, axial-vector, tensor hidden charmed tetraquark states and
axial-vector hidden bottom tetraquark states systematically  with the QCD sum rules \cite{Wang4430,WangHuang3900,Wang-tetraquark}, and assign the
$X(3872)$, $Z_c(3900/3885)$, $Z_c(4020/4025)$,  $Y(4140)$, $Z(4430)$, $Y(4660/4630)$ and $Z_b(10610/10650)$ to be tetraquark states tentatively,
\begin{eqnarray}
X(3872)&=&\frac{1}{2}\left( [cu]_{A}[\overline{c}\overline{u}]_{S} + [cd]_{A}[\overline{c}\overline{d}]_{S} +[cu]_{S}[\overline{c}\overline{u}]_{A}+[cd]_{S}[\overline{c}\overline{d}]_{A}\right) \,\,\,({\rm with}\,\,\,1^{++})\, , \nonumber \\
Z_c(3900/3885)&=&\frac{1}{\sqrt{2}}\left( [cu]_{A}[\overline{c}\overline{d}]_{S} - [cu]_{S}[\overline{c}\overline{d}]_{A}\right)\,\,\,({\rm with}\,\,\,1^{+-}) \, , \nonumber\\
Z_c(4020/4025)&=& [cu]_{A}[\overline{c}\overline{d}]_{A}   \,\,\,({\rm with}\,\,\,1^{+-}\,\,\,{\rm or}\,\,\,2^{++})\, ,\nonumber \\
Y(4140)&=& [cs]_{A}[\overline{c}\overline{s}]_{A}   \,\,\,({\rm with}\,\,\,2^{++})\, ,\nonumber \\
Z(4430)&=&\frac{1}{\sqrt{2}}\left( [cu]_{A}[\overline{c}\overline{d}]_{S} - [cu]_{S}[\overline{c}\overline{d}]_{A}\right)\,\,\,({\rm with}\,\,\,1^{+-}) \, , \nonumber\\
Y(4660/4630)&=&\frac{1}{\sqrt{2}}\left( [cs]_{A}[\overline{c}\overline{s}]_{P} - [cs]_{P}[\overline{c}\overline{s}]_{A}\right) \,\,\,({\rm with}\,\,\,1^{--})\, , \nonumber\\
Z_b(10610)&=&\frac{1}{\sqrt{2}}\left( [bu]_{A}[\overline{b}\overline{d}]_{S} - [bu]_{S}[\overline{b}\overline{d}]_{A}\right)\,\,\,({\rm with}\,\,\,1^{+-}) \, ,\nonumber \\
Z_b(10650)&=& [bu]_{A}[\overline{b}\overline{d}]_{A} \,\,\,({\rm with}\,\,\,1^{+-})\, ,
\end{eqnarray}
where the $[Qq]_{S}$, $[Qq]_{A}$ and $[Qq]_{P}$ denote the scalar, axial-vector and pseudoscalar diquarks  in the color antitriplet $\bar{3}$, $q=u,d,s$ and $Q=c,b$.

In the  scenario of  molecular states, we study the meson-meson type (or the $1\otimes1$ type) scalar, axial-vector and tensor  hadronic molecular states with the QCD sum rules \cite{WangHuang-molecule,Wang-molecule}, and assign
the $X(3872)$, $Z_c(3900/3885)$,   $Y(4140)$, $Z_c(4020/4025)$ and $Z_b(10610/10650)$ to be the molecular states tentatively,
\begin{eqnarray}
X(3872)&=&\frac{1}{\sqrt{2}}\left( D\overline{D}^{*} -  D^{*}\overline{D}\right) \,\,\,({\rm with}\,\,\,1^{++})\, , \nonumber \\
Z_c(3900/3885)&=&\frac{1}{\sqrt{2}}\left( D\overline{D}^{*} +  D^{*}\overline{D}\right)\,\,\,({\rm with}\,\,\,1^{+-}) \, , \nonumber\\
Z_c(4020/4025)&=& D^*\overline{D}^{*} \,\,\,({\rm with}\,\,\,1^{+-}\,\,\,{\rm or}\,\,\,2^{++})  \, ,\nonumber \\
Y(4140)&=&D_s^*\overline{D}_s^* \,\,\,({\rm with}\,\,\, 0^{++})\, , \nonumber\\
Z_b(10610)&=&\frac{1}{\sqrt{2}}\left( B\overline{B}^* + B^*\overline{B}\right)\,\,\,({\rm with}\,\,\,1^{+-}) \, ,\nonumber \\
Z_b(10650)&=&  B^{*}\overline{B}^{*} \,\,\,({\rm with}\,\,\,1^{+-}) \, .
\end{eqnarray}

There are more than one possibilities  in assigning the  $Z_c(3900/3885)$ and  $Z_c(4020/4025)$, we can assign them    either in the scenario of tetraquark states or molecular states.  The  $Z_c(3900)$, $Z_c(4200)$ and  $Z(4430)$ have the same quark constituents  $c\bar{c}u\bar{d}$ or $c\bar{c}d\bar{u}$ and analogous decays, see Eq.(1). Neither  scenario alone can accommodate the  $Z_c(3900)$, $Z_c(4200)$ and  $Z(4430)$ consistently, however, there are rooms to assign those $Z_c$ mesons consistently  in the two scenarios together, see section 4.
In Ref.\cite{WangHuang-molecule}, we observe that if we determine  the  energy scales of the QCD spectral densities with the same parameter ${\mathbb{M}}_c$, the $8\otimes8$ type molecule-like states have larger masses than the corresponding   $1\otimes1$ type molecular states. We obtain
 the masses $M_{Z_c(3900)}=3.89^{+0.09}_{-0.09}\,\rm{MeV}$ and $M_{Z_c({8\otimes8})}=4.10^{+0.09}_{-0.10}\,\rm{MeV}$ with  the QCD sum rules \cite{WangHuang-molecule}. The upper bound of the predicted  mass $M_{Z_c({8\otimes8})}=4.10^{+0.09}_{-0.10}\,\rm{MeV}$ is consistent with the experimental value $M_{Z_c(4200)} = 4196^{+31}_{-29}{}^{+17}_{-13} \,\rm{MeV}$ \cite{Zc4200exp}.
 Now, we assign the $Z_c(4200)$ to be the $8\otimes8$ type molecule-like state tentatively,
\begin{eqnarray}
Z_c(4200)&=&\frac{1}{\sqrt{2}}\left( \mathcal{D}\overline{\mathcal{D}}^{*} +  \mathcal{D}^{*}\overline{\mathcal{D}}\right)\,\,\,({\rm with}\,\,\,1^{+-}) \, ,
\end{eqnarray}
 study the mass and decay width with the QCD sum rules in details, and fit the effective mass ${\mathbb{M}}_c$ for the $8\otimes8$ type molecule-like states, where the $\mathcal{D}$ and $\mathcal{D}^*$ have the same quark constituents as the $D$ and $D^*$ respectively, but they are in the color $8$ representation.

The article is arranged as follows:  we derive the QCD sum rules for
the mass of the $8\otimes8$ type axial-vector molecule-like state  $Z_c(4200)$   in section 2; we derive the QCD sum rules for
the width of the $8\otimes8$ type axial-vector molecule-like state  $Z_c(4200)$   in section 3; we discuss the possible assignments of the
 $Z_c(3900)$, $Z_c(4200)$ and $Z(4430)$ as the $\bar{3}\otimes3$ type axial-vector tetraquark states  in section 4; section 5 is reserved for our conclusion.

\section{The mass of the $8\otimes8$ type axial-vector molecule-like state }
In the following, we write down  the two-point correlation function $\Pi_{\mu\nu}(p)$  in the QCD sum rules,
\begin{eqnarray}
\Pi_{\mu\nu}(p)&=&i\int d^4x e^{ip \cdot x} \langle0|T\left\{J_\mu(x)J_\nu^{\dagger}(0)\right\}|0\rangle \, , \\
J_\mu(x)&=&\frac{\bar{u}(x)i\gamma_5 \lambda^a c(x)\bar{c}(x)\gamma_\mu \lambda^a d(x)+\bar{u}(x)\gamma_\mu \lambda^a c(x)\bar{c}(x)i\gamma_5 \lambda^a d(x)}{\sqrt{2}} \, ,
\end{eqnarray}
where   the $\lambda^a$ is the Gell-Mann matrix in the color space. We construct the   $8\otimes8$ type  current $J_\mu(x)$ (see Ref.\cite{Wang-NPA}) to study the  molecule-like state  $Z_c(4200)$.
 Under charge conjugation transform $\widehat{C}$, the current $J_\mu(x)$ has the property,
\begin{eqnarray}
\widehat{C}J_{\mu}(x)\widehat{C}^{-1}&=&- J_\mu(x)\mid_{u {\leftrightarrow}d} \, .
\end{eqnarray}

We insert  a complete set of intermediate hadronic states with
the same quantum numbers as the current operator $J_\mu(x)$ into the
correlation function $\Pi_{\mu\nu}(p)$  to obtain the hadronic representation
\cite{SVZ79,Reinders85}, and isolate the ground state
contribution,
\begin{eqnarray}
\Pi_{\mu\nu}(p)&=&\frac{\lambda_{Z_c(4200)}^2}{M^2_{Z_c(4200)}-p^2}\left(-g_{\mu\nu} +\frac{p_\mu p_\nu}{p^2}\right) +\cdots  \, ,
\end{eqnarray}
where the pole residue  $\lambda_{Z_c(4200)}$ is defined by
\begin{eqnarray}
 \langle 0|J_\mu(0)|Z_c(4200)\rangle=\lambda_{Z_c(4200)}\, \varepsilon_\mu \, ,
\end{eqnarray}
the $\varepsilon_\mu$ is the polarization vector of the axial-vector meson $Z_c(4200)$.

We carry out the
operator product expansion to the vacuum condensates  up to dimension-10, and obtain the QCD spectral density through dispersion relation, then
 take the
quark-hadron duality and perform Borel transform  with respect to
the variable $P^2=-p^2$ to obtain  the following QCD sum rule,
\begin{eqnarray}
\lambda^2_{Z_c(4200)}\, \exp\left(-\frac{M^2_{Z_c(4200)}}{T^2}\right)= \int_{4m_c^2}^{s_0} ds\, \rho(s) \, \exp\left(-\frac{s}{T^2}\right) \, .
\end{eqnarray}
One can consult Ref.\cite{WangHuang-molecule} for the explicit expression of the QCD spectral density $\rho(s)$.
 We differentiate   Eq.(13) with respect to  $\tau=\frac{1}{T^2}$, then eliminate the
 pole residue  $\lambda_{Z_c(4200)}$ to obtain the QCD sum rule for the mass,
 \begin{eqnarray}
 M^2_{Z_c(4200)}= \frac{\int_{4m_c^2}^{s_0} ds\left(-\frac{d}{d \tau }\right)\rho(s)e^{-\tau s}}{\int_{4m_c^2}^{s_0} ds \rho(s)e^{-\tau s}}\, .
\end{eqnarray}

The   current $J_\mu(x)$  has non-vanishing couplings  with the scattering states  $D
D^\ast$, $J/\psi \pi$, $\eta_c \rho$, etc  \cite{PDG}.
 In the following, we
study the contributions of the  intermediate   meson-loops to the correlation function $\Pi_{\mu\nu}(p)$,
\begin{eqnarray}
\Pi_{\mu\nu}(p)&=&-\frac{\widehat{\lambda}_{Z_c}^{2}}{ p^2-\widehat{M}_{Z_c}^2}\widetilde{g}_{\mu\nu}(p)-\frac{\widehat{\lambda}_{Z_c}}{p^2-\widehat{M}_{Z_c}^2}\widetilde{g}_{\mu\alpha}(p)
 \Sigma_{DD^*}(p) \widetilde{g}^{\alpha\beta}(p) \widetilde{g}_{\beta\nu}(p)\frac{\widehat{\lambda}_{Z_c}}{p^2-\widehat{M}_{Z_c}^2} \nonumber \\
 &&-\frac{\widehat{\lambda}_{Z_c}}{p^2-\widehat{M}_{Z_c}^2}\widetilde{g}_{\mu\alpha}(p)
 \Sigma_{J/\psi\pi}(p) \widetilde{g}^{\alpha\beta}(p) \widetilde{g}_{\beta\nu}(p)\frac{\widehat{\lambda}_{Z_c}}{p^2-\widehat{M}_{Z_c}^2} \nonumber \\
 &&-\frac{\widehat{\lambda}_{Z_c}}{p^2-\widehat{M}_{Z_c}^2}\widetilde{g}_{\mu\alpha}(p)
 \Sigma_{\eta_c \rho}(p) \widetilde{g}^{\alpha\beta}(p) \widetilde{g}_{\beta\nu}(p)\frac{\widehat{\lambda}_{Z_c}}{p^2-\widehat{M}_{Z_c}^2}+\cdots \nonumber \\
 &=&-\frac{\widehat{\lambda}_{Z_c}^{2}}{ p^2-\widehat{M}_{Z_c}^2-\Sigma_{DD^*}(p)-\Sigma_{J/\psi\pi}(p)-\Sigma_{\eta_c \rho}(p)+\cdots}\widetilde{g}_{\mu\nu}(p)+\cdots \, , \end{eqnarray}
where
\begin{eqnarray}
\Sigma_{DD^*}(p)&=&i\int~{d^4q\over(2\pi)^4}\frac{G^2_{Z_cDD^*}}{\left[q^2-M_{D}^2\right]\left[ (p-q)^2-M_{D^*}^2\right]} \, ,\nonumber\\
\Sigma_{J/\psi \pi}(p)&=&i\int~{d^4q\over(2\pi)^4}\frac{G^2_{Z_cJ/\psi \pi}}{\left[q^2-M_{J/\psi}^2\right]\left[ (p-q)^2-M_{\pi}^2\right]} \, ,\nonumber\\
\Sigma_{\eta_c\rho}(p)&=&i\int~{d^4q\over(2\pi)^4}\frac{G^2_{Z_c\eta_c\rho}}{\left[q^2-M_{\eta_c}^2\right]\left[ (p-q)^2-M_{\rho}^2\right]}\, ,
\end{eqnarray}
$\widetilde{g}_{\mu\nu}(p)=-g_{\mu\nu}+\frac{p_{\mu}p_{\nu}}{p^2}$, the $G_{Z_c DD^*}$, $G_{Z_c J/\psi\pi}$, $G_{Z_c \eta_c \rho}$ are hadronic coupling constants, the $\widehat{\lambda}_{Z_c}$ and $\widehat{M}_{Z_c}$ are bare quantities to absorb the divergences in the self-energies $\Sigma_{DD^*}(p)$, $\Sigma_{J/\psi \pi}(p)$, $\Sigma_{\eta_c \rho}(p)$, etc.
The renormalized self-energies  contribute  a finite imaginary part to modify the dispersion relation,
\begin{eqnarray}
\Pi_{\mu\nu}(p) &=&-\frac{\lambda_{Z_c}^{2}}{ p^2-M_{Z_c}^2+i\sqrt{p^2}\Gamma(p^2)}\widetilde{g}_{\mu\nu}(p)+\cdots \, .
 \end{eqnarray}
 The present work (see section 3) indicates that
$\Gamma(Z_c^+(4200)\to J/\psi\pi^+)=24.6\,\rm{MeV}$, $ \Gamma(Z_c^+(4200)\to\eta_c\rho^+)=309.1 \,\rm{MeV}$ and $ \Gamma(Z_c^+(4200)\to(D\bar{D}^*/D^*\bar{D})^+)=0$   due to the special structure or interpolating current of the $Z_c(4200)$, the width originates dominantly  from the decays to the $\eta_c\rho$.
We can take into account the finite width effect by the following simple replacement of the hadronic spectral density,
\begin{eqnarray}
\lambda^2_{Z_c}\delta \left(s-M^2_{Z_c} \right) &\to& \lambda^2_{Z_c}\frac{1}{\pi}\frac{M_{Z_c}\Gamma_{Z_c}(s)}{(s-M_{Z_c}^2)^2+M_{Z_c}^2\Gamma_{Z_c}^2(s)}\, ,
\end{eqnarray}
where
\begin{eqnarray}
\Gamma_{Z_c}(s)&=&\Gamma_{Z_c} \frac{M_{Z_c}}{s}\sqrt{\frac{s-(M_{\eta_c}+M_{\rho})^2}{M^2_{Z_c}-(M_{\eta_c}+M_{\rho})^2}} \, .
\end{eqnarray}
Then the phenomenological sides of  the QCD sum rules in Eqs.(13-14) undergo the following changes,
\begin{eqnarray}
\lambda^2_{Z_c}\exp \left(-\frac{M^2_{Z_c}}{T^2} \right) &\to& \lambda^2_{Z_c}\int_{(M_{\eta_c}+M_{\rho})^2}^{s_0}ds\frac{1}{\pi}\frac{M_{Z_c}\Gamma_{Z_c}(s)}{(s-M_{Z_c}^2)^2+M_{Z_c}^2\Gamma_{Z_c}^2(s)}\exp \left(-\frac{s}{T^2} \right)\, , \nonumber\\
&=&0.76(0.79)\,\lambda^2_{Z_c}\exp \left(-\frac{M^2_{Z_c}}{T^2} \right)\, , \\
\lambda^2_{Z_c}M^2_{Z_c}\exp \left(-\frac{M^2_{Z_c}}{T^2} \right) &\to& \lambda^2_{Z_c}\int_{(M_{\eta_c}+M_{\rho})^2}^{s_0}ds\,s\,\frac{1}{\pi}\frac{M_{Z_c}\Gamma_{Z_c}(s)}{(s-M_{Z_c}^2)^2+M_{Z_c}^2\Gamma_{Z_c}^2(s)}\exp \left(-\frac{s}{T^2} \right)\, , \nonumber\\
&=&0.74(0.77)\,\lambda^2_{Z_c}M^2_{Z_c}\exp \left(-\frac{M^2_{Z_c}}{T^2} \right)\, ,
\end{eqnarray}
where we have used the central values of the input parameters, and $\Gamma_{Z_c}=370\,\rm{MeV}$, the values $0.79$ and $0.77$ in the bracket originate from the approximation $\Gamma(Z_c^+)\approx \Gamma(Z_c^+(4200)\to\eta_c\rho^+)$. So we can absorb the numerical factors  $0.76$ and $0.74$ into the pole residue $\lambda_{Z_c}$ safely, the intermediate   meson-loops cannot  affect  the mass $M_{Z_c(4200)}$ significantly,
 the zero width approximation in  the hadronic spectral density  works.

We take  the standard values of the vacuum condensates $\langle
\bar{q}q \rangle=-(0.24\pm 0.01\, \rm{GeV})^3$,   $\langle
\bar{q}g_s\sigma G q \rangle=m_0^2\langle \bar{q}q \rangle$,
$m_0^2=(0.8 \pm 0.1)\,\rm{GeV}^2$, $\langle \frac{\alpha_s
GG}{\pi}\rangle=(0.33\,\rm{GeV})^4 $    at the energy scale  $\mu=1\, \rm{GeV}$
\cite{SVZ79,Reinders85,Colangelo-Review}, and choose the $\overline{MS}$ mass $m_{c}(m_c)=(1.275\pm0.025)\,\rm{GeV}$  from the Particle Data Group \cite{PDG}.
Furthermore, we take into account the energy-scale dependence of  the input parameters,
\begin{eqnarray}
\langle\bar{q}q \rangle(\mu)&=&\langle\bar{q}q \rangle(Q)\left[\frac{\alpha_{s}(Q)}{\alpha_{s}(\mu)}\right]^{\frac{4}{9}}\, ,\nonumber\\
 \langle\bar{q}g_s \sigma Gq \rangle(\mu)&=&\langle\bar{q}g_s \sigma Gq \rangle(Q)\left[\frac{\alpha_{s}(Q)}{\alpha_{s}(\mu)}\right]^{\frac{2}{27}}\, ,\nonumber\\
m_c(\mu)&=&m_c(m_c)\left[\frac{\alpha_{s}(\mu)}{\alpha_{s}(m_c)}\right]^{\frac{12}{25}} \, ,\nonumber\\
\alpha_s(\mu)&=&\frac{1}{b_0t}\left[1-\frac{b_1}{b_0^2}\frac{\log t}{t} +\frac{b_1^2(\log^2{t}-\log{t}-1)+b_0b_2}{b_0^4t^2}\right]\, ,
\end{eqnarray}
  where $t=\log \frac{\mu^2}{\Lambda^2}$, $b_0=\frac{33-2n_f}{12\pi}$, $b_1=\frac{153-19n_f}{24\pi^2}$, $b_2=\frac{2857-\frac{5033}{9}n_f+\frac{325}{27}n_f^2}{128\pi^3}$,  $\Lambda=213\,\rm{MeV}$, $296\,\rm{MeV}$  and  $339\,\rm{MeV}$ for the flavors  $n_f=5$, $4$ and $3$, respectively  \cite{PDG}.
We  tentatively take the continuum threshold parameter  as $\sqrt{s_0}=(4.7\pm 0.1)\,\rm{GeV}^2$, i.e. $\sqrt{s_0}=M_{Z_c(4200)}+(0.4-0.6)\,\rm{GeV}$,  and search for the optimal Borel parameter to satisfy the two criteria (pole dominance and convergence of the operator product
expansion) of the QCD sum rules.

In Fig.1,  we plot the mass of the $Z_c(4200)$    with variations of the  Borel parameters $T^2$ and energy scales $\mu$ for the
threshold parameter $\sqrt{s_0}=4.7\,\rm{GeV}$. From the figure, we can see that the masses decrease monotonously
with increase of the energy scales. We can reproduce the experimental value $M_{Z_c(4200)} = 4196^{+31}_{-29}{}^{+17}_{-13} \,\rm{MeV}$   \cite{Zc4200exp} approximately  at the energy scale $\mu=(1.3-1.5)\,\rm{GeV}$. In this article, we take the energy scale $\mu=1.4\,\rm{GeV}$. In the Borel window $T^2=(3.0-3.4)\,\rm{GeV}^2$, the pole contribution is about $(43-64)\%$, it is reliable to extract the ground state mass.
\begin{figure}
 \centering
 \includegraphics[totalheight=8cm,width=12cm]{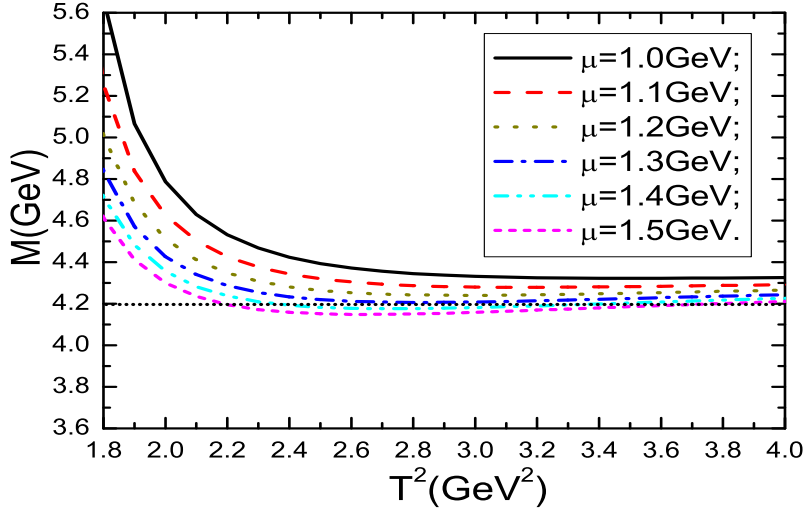}
    \caption{The mass  of the $Z_c(4200)$ with variations of the Borel parameters $T^2$ and energy scales $\mu$, where  the horizontal line denotes  the experimental value. }
\end{figure}

In Fig.2,  we plot the contributions of different terms in the
operator product expansion with variations of the Borel parameters  $T^2$ for the threshold parameter $\sqrt{s_0}=4.7\,\rm{GeV}$ and energy scale $\mu=1.4\,\rm{GeV}$. In the Borel window $T^2=(3.0-3.4)\,\rm{GeV}^2$, the  $D_3\gg D_0\approx |D_5|\gg D_6\gg |D_8|$, and the $D_4$, $D_7$ and $D_{10}$ play a less important role,
where the $D_i$ with $i=0,\,3,\,4,\,5,\,6,\,7,\,8,\,10$ denote the contributions of the vacuum condensates of dimensions $D=i$, and the total contributions are normalized to be $1$. The operator product expansion is well convergent.

\begin{figure}
 \centering
 \includegraphics[totalheight=8cm,width=12cm]{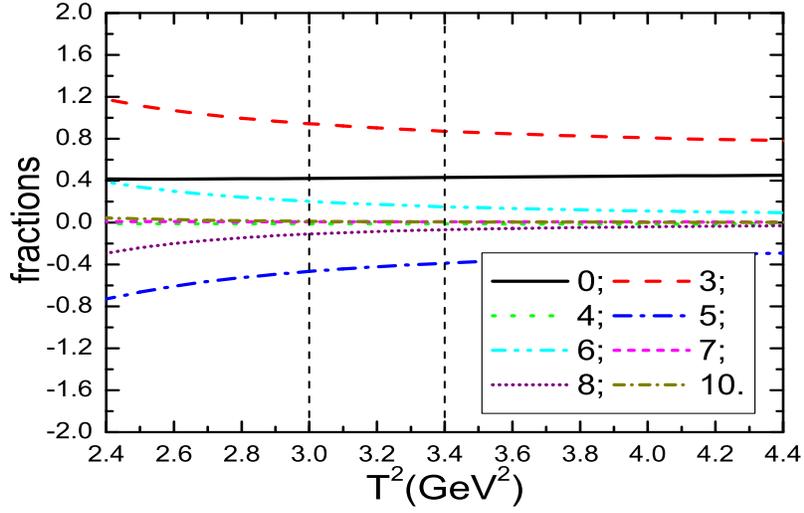}
    \caption{The contributions of different terms in the operator product expansion  with variations of the
  Borel parameter $T^2$, where the $0$, $3$, $4$, $5$, $6$, $7$, $8$ and $10$   denotes the dimensions of the vacuum condensates. }
\end{figure}

\begin{figure}
 \centering
 \includegraphics[totalheight=8cm,width=12cm]{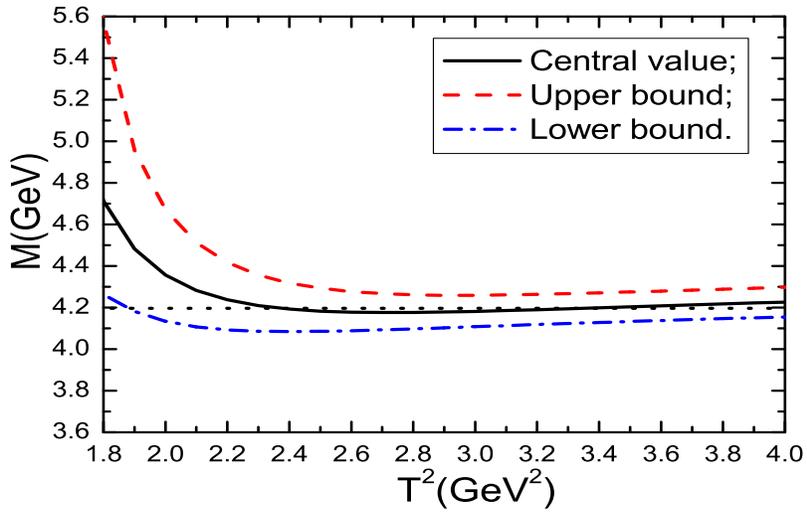}
    \caption{The mass  of the $Z_c(4200)$ with variations of the Borel parameter $T^2$,  where  the horizontal line denotes  the experimental value.  }
\end{figure}

\begin{figure}
 \centering
 \includegraphics[totalheight=8cm,width=12cm]{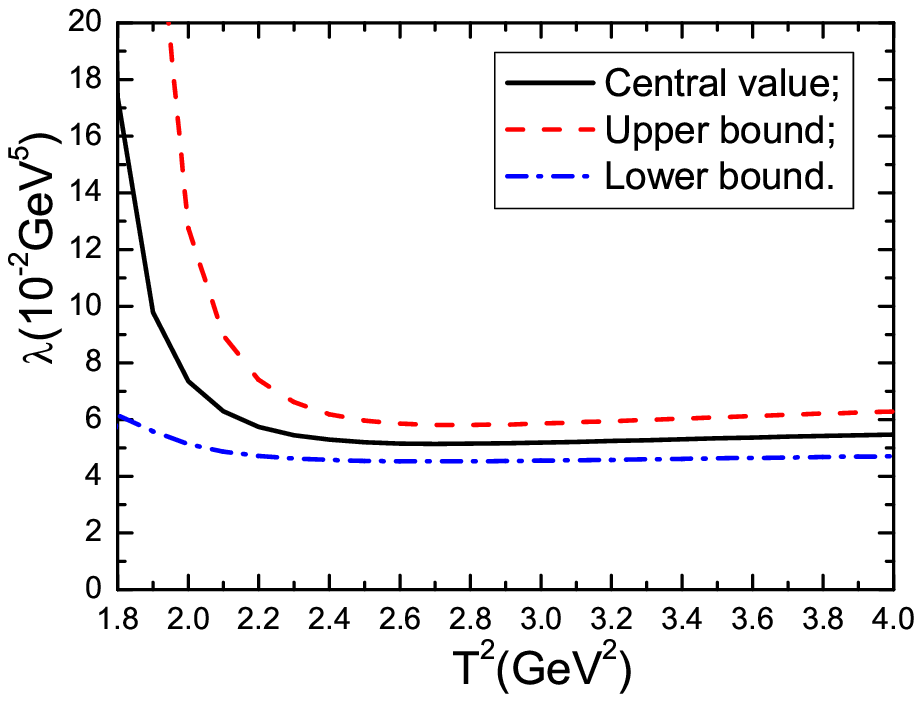}
    \caption{The pole residue  of the $Z_c(4200)$ with variations of the Borel parameter $T^2$.  }
\end{figure}

We take  into account all uncertainties of the input parameters,
and obtain the values of the mass and pole residue of
 the    $Z_c(4200)$, which are  shown explicitly in Figs.3-4,
\begin{eqnarray}
M_{Z_c(4200)}&=&4.19 \pm 0.08\,\rm{GeV} \, ,  \nonumber\\
\lambda_{Z_c(4200)}&=&(5.25\pm0.71)\times 10^{-2}\,\rm{GeV}^5 \,   .
\end{eqnarray}
The predicted mass $M_{Z_c(4200)}=4.19 \pm 0.08\,\rm{GeV}$  is consistent with the experimental value $M_{Z_c(4200)} = 4196^{+31}_{-29}{}^{+17}_{-13} \,\rm{MeV}$ within uncertainties \cite{Zc4200exp}.  The QCD sum rules favor assigning the $Z_c(4200)$ to be  the $8\otimes8$ type $ \mathcal{D}\overline{\mathcal{D}}^{*} +  \mathcal{D}^{*}\overline{\mathcal{D}}$ molecule-like  state.
Now we can obtain the parameter ${\mathbb{M}}_c=1.98\,\rm{GeV}$ for the $8\otimes8$ type   molecule-like  states according to the energy scale formula  $\mu=\sqrt{M^2_{X/Y/Z}-(2{\mathbb{M}}_c)^2}$.

\section{The width of the $8\otimes8$ type axial-vector molecule-like state }

We can study the strong decays $Z_c^\pm(4200)\to J/\psi\pi^{\pm}$, $\eta_c\rho^{\pm}$ and $(D\bar{D}^*)^\pm$ (or $(D^*\bar{D})^\pm$) with the following three-point correlation functions
$\Pi_{\alpha\mu}^{1}(p,q)$, $\Pi_{\alpha\mu}^{2}(p,q)$ and $\Pi_{\alpha\mu}^{3}(p,q)$, respectively,
\begin{eqnarray}
\Pi_{\alpha\mu}^{1}(p,q)&=&i^2\int d^4xd^4y e^{ipx}e^{iqy}\langle 0|T\left\{J_\alpha^{J/\psi}(x)J_5^{\pi}(y)J_{\mu}(0)\right\}|0\rangle\, ,   \\
\Pi_{\alpha\mu}^{2}(p,q)&=&i^2\int d^4xd^4y e^{ipx}e^{iqy}\langle 0|T\left\{J_5^{\eta_c}(x)J_\alpha^{\rho}(y)J_{\mu}(0)\right\}|0\rangle \, , \\
\Pi_{\alpha\mu}^{3}(p,q)&=&i^2\int d^4xd^4y e^{ipx}e^{iqy}\langle 0|T\left\{J_\alpha^{D^*}(x)J_5^{D}(y)J_{\mu}(0)\right\}|0\rangle \, ,
\end{eqnarray}
where the currents
\begin{eqnarray}
J_\alpha^{J/\psi}(x)&=&\bar{c}(x)\gamma_\alpha c(x) \, ,\nonumber \\
J_5^{\pi}(y)&=&\bar{u}(y)i\gamma_5 d(y) \, ,  \\
J_5^{\eta_c}(x)&=&\bar{c}(x)i\gamma_5 c(x) \, ,\nonumber \\
J_\alpha^{\rho}(y)&=&\bar{u}(y)\gamma_\alpha d(y) \, ,  \\
J_\alpha^{D^*}(x)&=&\bar{c}(x)\gamma_\alpha d(x) \, ,\nonumber \\
J_5^{D}(y)&=&\bar{u}(y)i\gamma_5 c(y) \, ,
\end{eqnarray}
interpolate the mesons $J/\psi$, $\pi$, $\eta_c$, $\rho$, $D^*$ and $D$, respectively. At the leading order ${\mathcal{O}}(\alpha_s)$, $\Pi_{\alpha\mu}^{3}(p,q)=0$ at the QCD side according to the structure of the $SU(3)$ color group.

We insert  a complete set of intermediate hadronic states with
the same quantum numbers as the current operators into the three-point
correlation functions $\Pi_{\alpha\mu}^{1}(p,q)$, $\Pi_{\alpha\mu}^{2}(p,q)$ and  isolate the ground state
contributions to obtain the following results,
\begin{eqnarray}
\Pi_{\alpha\mu}^{1}(p,q)&=& \frac{f_{\pi}M_{\pi}^2f_{J/\psi}M_{J/\psi}\lambda_{Z_c}G_{Z_cJ/\psi \pi}}{m_u+m_d} \frac{1}{(M_{Z_c}^2-p^{\prime2})(M_{J/\psi}^2-p^2)(M_{\pi}^2-q^2)} \left(-g_{\alpha\beta}+\frac{p_{\alpha}p_{\beta}}{p^2} \right) \nonumber\\
&&\left(-g_{\mu}^{\beta}+\frac{p^{\prime}_{\mu}p^{\prime\beta}}{p^{\prime2}} \right)+\cdots  \nonumber\\
&=&\left\{ \frac{f_{\pi}M_{\pi}^2f_{J/\psi}M_{J/\psi}\lambda_{Z_c}G_{Z_cJ/\psi \pi}}{m_u+m_d} \frac{1}{(M_{Z_c}^2-p^{\prime2})(M_{J/\psi}^2-p^2)(M_{\pi}^2-q^2)}\right.\nonumber\\
&&+ \frac{1}{(M_{Z_c}^2-p^{\prime2})(M_{J/\psi}^2-p^2)} \int_{s^0_\pi}^\infty dt\frac{\rho_{Z_c\pi}(p^2,t,p^{\prime 2})}{t-q^2}\nonumber\\
&&\left.+ \frac{1}{(M_{Z_c}^2-p^{\prime2})(M_{\pi}^2-q^2)} \int_{s^0_{J/\psi}}^\infty dt\frac{\rho_{Z_cJ/\psi}(t,q^2,p^{\prime 2})}{t-p^2}+\cdots\right\}\left(g_{\alpha\mu}+\cdots\right) +\cdots\nonumber\\
&=&\left\{ \frac{f_{\pi}M_{\pi}^2f_{J/\psi}M_{J/\psi}\lambda_{Z_c}G_{Z_cJ/\psi \pi}}{m_u+m_d} \frac{1}{(M_{Z_c}^2-p^{\prime2})(M_{J/\psi}^2-p^2)(M_{\pi}^2-q^2)}\right.\nonumber\\
&&\left.+ \frac{C_{Z_c\pi}}{(M_{Z_c}^2-p^{\prime2})(M_{J/\psi}^2-p^2)}
 + \frac{C_{Z_cJ/\psi}}{(M_{Z_c}^2-p^{\prime2})(M_{\pi}^2-q^2)} +\cdots\right\}\left(g_{\alpha\mu}+\cdots\right)+\cdots \, , \nonumber\\
\end{eqnarray}

\begin{eqnarray}
\Pi_{\alpha\mu}^{2}(p,q)&=& \frac{f_{\eta_c}M_{\eta_c}^2f_{\rho}M_{\rho}\lambda_{Z_c}G_{Z_c\eta_c \rho}}{2m_c} \frac{1}{(M_{Z_c}^2-p^{\prime2})(M_{\eta_c}^2-p^2)(M_{\rho}^2-q^2)} \left(-g_{\alpha\beta}+\frac{q_{\alpha}q_{\beta}}{q^2} \right) \nonumber\\
&&\left(-g_{\mu}^{\beta}+\frac{p^{\prime}_{\mu}p^{\prime\beta}}{p^{\prime2}} \right)+\cdots  \nonumber\\
&=&\left\{ \frac{f_{\eta_c}M_{\eta_c}^2f_{\rho}M_{\rho}\lambda_{Z_c}G_{Z_c\eta_c \rho}}{2m_c} \frac{1}{(M_{Z_c}^2-p^{\prime2})(M_{\eta_c}^2-p^2)(M_{\rho}^2-q^2)}\right.\nonumber\\
&&+  \frac{1}{(M_{Z_c}^2-p^{\prime2})(M_{\eta_c}^2-p^2)}\int_{s^0_\rho}^\infty dt \frac{\rho_{Z_c\rho}(p^2,t,p^{\prime 2})}{t-q^2} \nonumber\\
&&\left.+ \frac{1}{(M_{Z_c}^2-p^{\prime2})(M_{\rho}^2-q^2)} \int_{s^0_{\eta_c}}^\infty dt\frac{\rho_{Z_c\eta_c}(t,q^2,p^{\prime 2})}{t-p^2}+\cdots\right\}\left(g_{\alpha\mu}+\cdots\right) +\cdots\nonumber\\
&=&\left\{ \frac{f_{\eta_c}M_{\eta_c}^2f_{\rho}M_{\rho}\lambda_{Z_c}G_{Z_c\eta_c \rho}}{2m_c} \frac{1}{(M_{Z_c}^2-p^{\prime2})(M_{\eta_c}^2-p^2)(M_{\rho}^2-q^2)}\right.\nonumber\\
&&\left.+  \frac{C_{Z_c\rho}}{(M_{Z_c}^2-p^{\prime2})(M_{\eta_c}^2-p^2)}+ \frac{C_{Z_c\eta_c}}{(M_{Z_c}^2-p^{\prime2})(M_{\rho}^2-q^2)} +\cdots\right\}\left(g_{\alpha\mu}+\cdots\right)+\cdots \, , \nonumber\\
\end{eqnarray}
where $p^\prime=p+q$, the $f_{J/\psi}$, $f_{\eta_c}$, $f_{\rho}$ and $f_{\pi}$ are the decay constants of the mesons  $J/\psi$, $\eta_c$, $\rho$ and $\pi$, respectively, the $G_{Z_cJ/\psi\pi}$ and $G_{Z_c\eta_c\rho}$ are the hadronic coupling constants.  In this article, we choose the tensor $g_{\alpha\mu}$  to study the  coupling constants $G_{Z_cJ/\psi\pi}$ and $G_{Z_c\eta_c\rho}$.

In the following, we write down the definitions,
\begin{eqnarray}
\langle0|J_{\alpha}^{J/\psi}(0)|J/\psi(p)\rangle&=&f_{J/\psi}M_{J/\psi}\xi_\alpha \,\, , \nonumber \\
\langle0|J_{\alpha}^{\rho}(0)|\rho(q)\rangle&=&f_{\rho}M_{\rho}\varepsilon_\alpha \,\, , \nonumber \\
\langle0|J_{5}^{\eta_c}(0)|\eta_c(p)\rangle&=&\frac{f_{\eta_c}M_{\eta_c}^2}{2m_c} \,\, , \nonumber \\
\langle0|J_{5}^{\pi}(0)|\pi(q)\rangle&=&\frac{f_{\pi}M_{\pi}^2}{m_u+m_d} \,\, ,  \\
\langle J/\psi(p)\pi(q)|Z_c(p^{\prime})\rangle&=&i\xi^*(p)\cdot\zeta(p^{\prime}) G_{Z_cJ/\psi\pi} \, , \nonumber\\
\langle\eta_c(p)\rho(q)|Z_c(p^{\prime})\rangle&=&i\varepsilon^*(q)\cdot\zeta(p^{\prime}) G_{Z_c\eta_c\rho}  \, ,
\end{eqnarray}
the $\xi$, $\zeta$ and $\varepsilon$ are polarization vectors of the $J/\psi$, $Z_c(4200)$ and $\rho$, respectively.
The four unknown functions $\rho_{Z_c\pi}(p^2,t,p^{\prime 2})$, $\rho_{Z_cJ/\psi}(t,q^2,p^{\prime 2})$, $\rho_{Z_c\rho}(p^2,t,p^{\prime 2})$ and $\rho_{Z_c\eta_c}(t,q^2,p^{\prime 2})$ have complex dependence on the transitions
between the ground states and the high resonances  or the continuum states. We introduce the notations $C_{Z_c\pi}$, $C_{Z_cJ/\psi}$, $C_{Z_c\rho}$ and $C_{Z_c\eta_c}$ to parameterize the net effects,
\begin{eqnarray}
C_{Z_c\pi}&=&\int_{s^0_\pi}^\infty dt\frac{ \rho_{Z_c\pi}(p^2,t,p^{\prime 2})}{t-q^2}\, ,\nonumber\\
C_{Z_cJ/\psi}&=&\int_{s^0_{J/\psi}}^\infty dt\frac{\rho_{Z_cJ/\psi}(t,q^2,p^{\prime 2})}{t-p^2}\, ,\nonumber\\
C_{Z_c\rho}&=&\int_{s^0_{\rho}}^\infty dt \frac{\rho_{Z_c\rho}(p^2,t,p^{\prime 2})}{t-q^2}\, ,\nonumber\\
C_{Z_c\eta_c}&=&\int_{s^0_{\eta_c}}^\infty dt \frac{ \rho_{Z_c\eta_c}(t,q^2,p^{\prime 2})}{t-p^2}\, .
\end{eqnarray}
In numerical calculations,   we smear  the dependencies of the  $C_{Z_c\pi}$, $C_{Z_cJ/\psi}$, $C_{Z_c\rho}$ and $C_{Z_c\eta_c}$ on the variables $p^2,\,p^{\prime 2},\,q^2$, take the $C_{Z_c\pi}$, $C_{Z_cJ/\psi}$, $C_{Z_c\rho}$ and $C_{Z_c\eta_c}$ as free parameters, and choose the suitable values  to
eliminate the contaminations to obtain the stable sum rules with the variations of
the Borel parameters \cite{Ioffe-84}.

We carry out the operator product expansion up to the vacuum condensates of dimension 5 and neglect the tiny contribution of the gluon condensate, one can see Fig.2 as an example. In the QCD sum rules involving  the tetraquark states, if   there exist contributions from the perturbative terms or quark condensate terms, or mixed   condensate terms, then the gluon condensates play a minor important role. The leading-order perturbative terms, the quark condensates, the mixed condensates and the gluon condensates  are    the vacuum expectations
of the operators of the order    $\mathcal{O}( \alpha_s^{0})$, $\mathcal{O}(\alpha_s^0)$, $\mathcal{O}( \alpha_s^{1/2})$ and $\mathcal{O}( \alpha_s)$, respectively, so the gluon condensates can be neglected approximately. Furthermore, the double Borel transformed QCD sum rules converge  much faster than the single Borel transformed QCD sum rules at the operator product expansion side.
We obtain the QCD spectral densities through dispersion relation, take the quark-hadron  duality below the continuum thresholds, then set $p^{\prime2}=p^2$ and take  the double Borel transforms with respect to the variable   $P^2=-p^2 $ and $Q^2=-q^2$ respectively to obtain the following QCD sum rules,
\begin{eqnarray}
&&\frac{f_{\pi}M_{\pi}^2f_{J/\psi}M_{J/\psi}\lambda_{Z_c}G_{Z_cJ/\psi \pi}}{m_u+m_d}\frac{1}{M_{Z_c}^2-M_{J/\psi}^2} \left[ \exp\left(-\frac{M_{J/\psi}^2}{T_1^2} \right)-\exp\left(-\frac{M_{Z_c}^2}{T_1^2} \right)\right]\exp\left(-\frac{M_{\pi}^2}{T_2^2} \right) \nonumber\\
&&+C_{Z_cJ/\psi} \exp\left(-\frac{M_{Z_c}^2}{T_1^2} -\frac{M_{\pi}^2}{T_2^2} \right)=\frac{1}{12\sqrt{2}\pi^4}\int_{4m_c^2}^{s^0_{Z_c}} ds \int_{0}^{s^0_{\pi}} du  u\left(s+2m_c^2\right)\sqrt{1-\frac{4m_c^2}{s}}\nonumber\\
&&\exp\left(-\frac{s}{T_1^2} -\frac{u}{T_2^2} \right)-\frac{m_c\langle\bar{q}g_s\sigma Gq\rangle}{6\sqrt{2}\pi^2} \int_{4m_c^2}^{s^0_{Z_c}} ds \sqrt{1-\frac{4m_c^2}{s}}\exp\left(-\frac{s}{T_1^2}  \right)\, ,
\end{eqnarray}
\begin{eqnarray}
&&\frac{f_{\eta_c}M_{\eta_c}^2f_{\rho}M_{\rho}\lambda_{Z_c}G_{Z_c\eta_c \rho}}{2m_c } \frac{1}{M_{Z_c}^2-M_{\eta_c}^2}\left[ \exp\left(-\frac{M_{\eta_c}^2}{T_1^2} \right)-\exp\left(-\frac{M_{Z_c}^2}{T_1^2} \right)\right]\exp\left(-\frac{M_{\rho}^2}{T_2^2} \right) \nonumber\\
&&+C_{Z_c\eta_c} \exp\left(-\frac{M_{Z_c}^2}{T_1^2} -\frac{M_{\rho}^2}{T_2^2} \right)=\frac{1}{12\sqrt{2}\pi^4}\int_{4m_c^2}^{s^0_{Z_c}} ds \int_{0}^{s^0_{\rho}} du  us\sqrt{1-\frac{4m_c^2}{s}}\exp\left(-\frac{s}{T_1^2} -\frac{u}{T_2^2} \right)\nonumber\\
&&+\frac{m_c\langle\bar{q}g_s\sigma Gq\rangle}{18\sqrt{2}\pi^2} \int_{4m_c^2}^{s^0_{Z_c}} ds \sqrt{1-\frac{4m_c^2}{s}}\exp\left(-\frac{s}{T_1^2}  \right)\, ,
\end{eqnarray}
where the $s^0_{Z_c}$, $s^0_{\pi}$ and $s^0_{\rho}$ are the continuum threshold parameters for the $Z_c(4200)$, $\pi$ and $\rho$, respectively.

The hadronic parameters are taken as $M_{\pi}=0.13957\,\rm{GeV}$,  $M_{\rho}=0.77526\,\rm{GeV}$,
$M_{J/\psi}=3.0969\,\rm{GeV}$, $M_{\eta_c}=2.9836\,\rm{GeV}$ \cite{PDG},  $f_{\pi}=0.130\,\rm{GeV}$, $f_{\rho}=0.215\,\rm{GeV}$ \cite{Colangelo-Review},
$f_{J/\psi}=0.418 \,\rm{GeV}$, $f_{\eta_c}=0.387 \,\rm{GeV}$  \cite{Becirevic}, $\sqrt{s^0_{\pi}}=0.85\,\rm{GeV}$, $\sqrt{s^0_{\rho}}=1.3\,\rm{GeV}$ \cite{Colangelo-Review}, $\sqrt{s^0_{Z_c}}=4.7\,\rm{GeV}$,   $\lambda_{Z_c}=5.25\times 10^{-2}\,\rm{GeV}^5$, $T_1^2=(3.0-3.4)\,\rm{GeV}^2$ (present work), $T_2^2=(0.8-1.2)\,\rm{GeV}^2$ \cite{Colangelo-Review}, $f_{\pi}M^2_{\pi}/(m_u+m_d)=-2\langle \bar{q}q\rangle/f_{\pi}$ from the Gell-Mann-Oakes-Renner relation.
The unknown parameters are chosen as $C_{Z_cJ/\psi}=0.01\,\rm{GeV}^6 $ and $C_{Z_c\eta_c}=0.09\,\rm{GeV}^6 $  to obtain  platforms in the Borel windows $T_1^2=(3.0-3.4)\,\rm{GeV}^2$. The parameters at the QCD side are chosen as the same in the two-point QCD sum rules for the $Z_c(4200)$.
Then it is easy to obtain the values of the hadronic coupling constants,
\begin{eqnarray}
G_{Z_cJ/\psi \pi} &=&3.34\pm0.07\pm0.25\,\rm{GeV}\, , \nonumber\\
G_{Z_c\eta_c\rho}&=&11.31\pm0.07\pm1.06\,\rm{GeV}\, ,
\end{eqnarray}
where the uncertainties originate from the Borel parameters $T_1^2$ and $T_2^2$, respectively, see Figs.5-6. As the largest uncertainties originate from the Borel parameter $T_2^2$, we neglect the uncertainties of the parameters  other than the Borel parameters. The uncertainties of the $G_{Z_cJ/\psi \pi}$ and $G_{Z_c\eta_c\rho}$ lead to the uncertainties $\delta\Gamma(Z_c^+(4200)\to J/\psi\pi^+)/\Gamma(Z_c^+(4200)\to J /\psi\pi^+)=2\delta G_{Z_cJ/\psi \pi}/G_{Z_cJ/\psi \pi}\approx 8\%$ and $\delta\Gamma(Z_c^+(4200)\to \eta_c\rho^+)/\Gamma(Z_c^+(4200)\to \eta_c\rho^+)=2\delta G_{Z_c\eta_c \rho}/G_{Z_c\eta_c \rho}\approx 9\%$.

\begin{figure}
 \centering
 \includegraphics[totalheight=8cm,width=12cm]{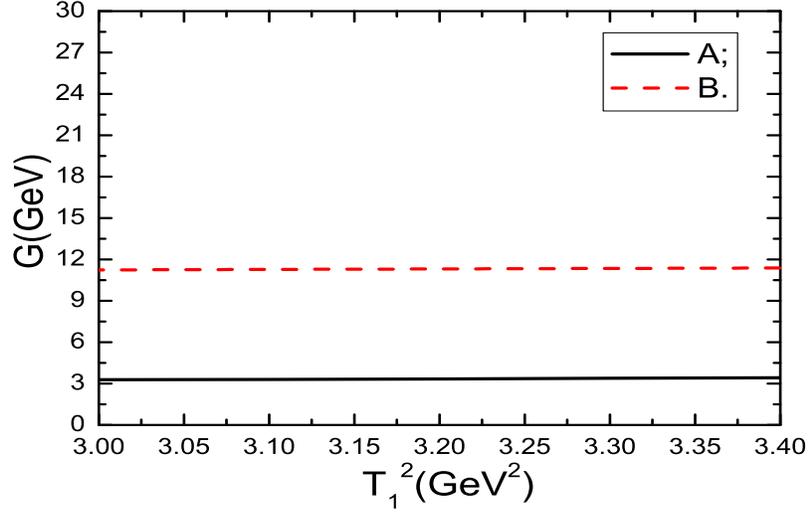}
    \caption{The coupling constants $G_{Z_cJ/\psi\pi}$ ($A$) and  $G_{Z_c\eta_c\rho}$ ($B$)  with variations of the Borel parameters $T_1^2$.  }
\end{figure}

\begin{figure}
 \centering
 \includegraphics[totalheight=8cm,width=12cm]{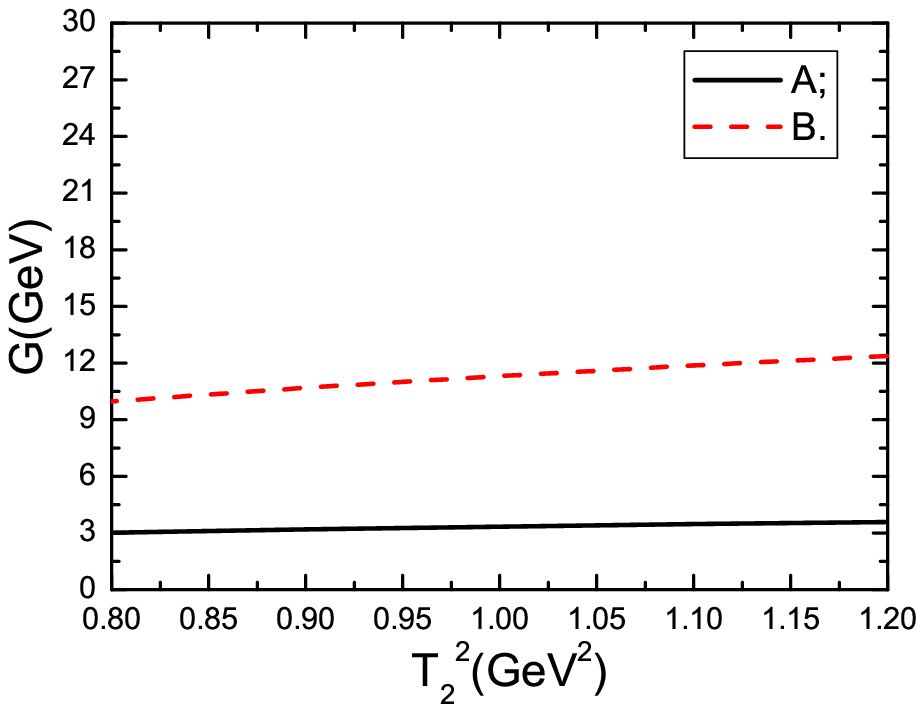}
    \caption{The coupling constants $G_{Z_cJ/\psi\pi}$ ($A$) and  $G_{Z_c\eta_c\rho}$ ($B$)  with variations of the Borel parameters $T_2^2$.  }
\end{figure}

The central values of the   decay widths are
\begin{eqnarray}
\Gamma(Z_c^+(4200)\to J/\psi\pi^+)&=& \frac{p\left(M_{Z_c},M_{J/\psi},M_{\pi}\right)}{24\pi M_{Z_c}^2}G_{Z_cJ/\psi\pi}^2\left[3+\frac{p\left(M_{Z_c},M_{J/\psi},M_{\pi}\right)^2}{M_{J/\psi}^2} \right]\nonumber\\
&=&24.6\,\rm{MeV}   \, ,\nonumber\\
\Gamma(Z_c^+(4200)\to\eta_c\rho^+)&=& \frac{p\left(M_{Z_c},M_{\eta_c},M_{\rho}\right)}{24\pi M_{Z_c}^2}G_{Z_c\eta_c\rho}^2\left[3+\frac{p\left(M_{Z_c},M_{\eta_c},M_{\rho}\right)^2}{M_{\rho}^2} \right]\nonumber\\
&=&309.1 \,\rm{MeV}  \, ,
\end{eqnarray}
where $p(a,b,c)=\frac{\sqrt{[a^2-(b+c)^2][a^2-(b-c)^2]}}{2a}$.	If we saturate the width of the $Z_c(4200)$ with the strong decays to $J/\psi\pi$ and $\eta_c\rho$, then
$\Gamma_{Z_c(4200)}\approx 334\,\rm{MeV}$, which is consistent with the experimental value $\Gamma_{Z_c(4200)} = 370^{+70}_{-70}{}^{+70}_{-132}\,\rm{MeV}$ from the Belle collaboration  \cite{Zc4200exp}, the present calculations support  assigning the $Z_c(4200)$ to be  the $8\otimes8$ type axial-vector  molecule-like state. Due to the special structure of the $Z_c(4200)$, the decays to the final states $D\bar{D}^*$ and $D^*\bar{D}$ can only take place through re-scattering mechanism $Z_c(4200)\to J/\psi\pi,\,\eta_c\rho\to D\bar{D}^*,\,D^*\bar{D}$, the decay widths $\Gamma(Z_c(4200)\to  D\bar{D}^*,\,D^*\bar{D})$ are expected to be small.

\section{The masses of the $\bar{3}\otimes3$ type axial-vector tetraquark  states }
In the following, we write down  the two-point correlation function $\overline{\Pi}_{\mu\nu}(p)$  in the QCD sum rules,
\begin{eqnarray}
\overline{\Pi}_{\mu\nu}(p)&=&i\int d^4x e^{ip \cdot x} \langle0|T\left\{\eta_\mu(x)\eta_\nu^{\dagger}(0)\right\}|0\rangle \, , \\
\eta_\mu(x)&=&\frac{\epsilon^{ijk}\epsilon^{imn}}{\sqrt{2}}\left\{u^j(x)C\gamma_5c^k(x) \bar{d}^m(x)\gamma_\mu C \bar{c}^n(x)-u^j(x)C\gamma_\mu c^k(x)\bar{d}^m(x)\gamma_5C \bar{c}^n(x) \right\} \, ,
\end{eqnarray}
 the $i$, $j$, $k$, $m$, $n$ are color indexes, the $C$ is the charge conjugation matrix. We choose  the   current $\eta_\mu(x)$ to interpolate the
   $\bar{3}\otimes 3$ type axial-vector  tetraquark states.  Under charge conjugation transform $\widehat{C}$,
  the current $\eta_\mu(x)$ has the property,
\begin{eqnarray}
\widehat{C}\eta_{\mu}(x)\widehat{C}^{-1}&=&- \eta_\mu(x)\mid_{u \leftrightarrow d} \, .
\end{eqnarray}
We carry out the operator product expansion up to the vacuum condensates of dimension 10 and obtain the correlation function at the QCD side,
\begin{eqnarray}
\overline{\Pi}_{\mu\nu}(p)&=&\int_{4m_c^2}^\infty ds \frac{\rho(s)}{s-p^2} \left(-g_{\mu\nu}+\frac{p_{\mu} p_{\nu}}{p^2} \right)+\overline{\Pi}_0(p^2) \frac{p_{\mu} p_{\nu}}{p^2} \, ,
\end{eqnarray}
the expression of the QCD spectral density $\rho(s)$ is shown explicitly in Ref.\cite{WangHuang3900}, the component $\overline{\Pi}_0(p^2)$ is irrelevant in the present analysis.

In case I, the  $Z_c(3900)$ and $Z(4430)$  are the ground state and the first radial excited state of the  $\bar{3}\otimes 3$ type axial-vector tetraquark states, respectively, the $Z_c(4200)$ is not the $\bar{3}\otimes 3$ type axial-vector tetraquark state, then the current couples potentially to the $Z_c(3900)$ and $Z(4430)$, $\langle 0|\eta_\mu(0)|Z_c(3900)/Z(4430)\rangle=\lambda_{Z_c(3900)/Z(4430)}\varepsilon_\mu$, where the $\varepsilon_\mu$ are  the polarization vectors of the $Z_c(3900)$ and $Z(4430)$. Now we retain the ground state and the first radial excited state and write down the QCD sum rule \cite{Wang4430},
\begin{eqnarray}
\lambda_{Z_c(3900)}^2\, \exp\left(-\frac{M_{Z_c(3900)}^2}{T^2}\right)+\lambda_{Z(4430)}^2\,
 \exp\left(-\frac{M_{Z(4430)}^2}{T^2}\right) &=& \int_{4m_c^2}^{s_0} ds\, \rho(s) \, \exp\left(-\frac{s}{T^2}\right) \, . \nonumber\\
\end{eqnarray}
 We differentiate   Eq.(43) with respect to  $\tau=\frac{1}{T^2}$  and obtain three additional QCD sum rules,
 \begin{eqnarray}
&&\lambda_{Z_c(3900)}^2M_{Z_c(3900)}^2\, \exp\left(-\frac{M_{Z_c(3900)}^2}{T^2}\right)+\lambda_{Z(4430)}^2M_{Z(4430)}^2\,
 \exp\left(-\frac{M_{Z(4430)}^2}{T^2}\right) \nonumber\\
 && =\int_{4m_c^2}^{s_0} ds\, s\rho(s) \, \exp\left(-\frac{s}{T^2}\right) \, ,
\end{eqnarray}
\begin{eqnarray}
&&\lambda_{Z_c(3900)}^2M_{Z_c(3900)}^4\, \exp\left(-\frac{M_{Z_c(3900)}^2}{T^2}\right)+\lambda_{Z(4430)}^2M_{Z(4430)}^4\,
 \exp\left(-\frac{M_{Z(4430)}^2}{T^2}\right) \nonumber\\
 && =\int_{4m_c^2}^{s_0} ds\, s^2\rho(s) \, \exp\left(-\frac{s}{T^2}\right) \, ,
\end{eqnarray}
\begin{eqnarray}
&&\lambda_{Z_c(3900)}^2M_{Z_c(3900)}^6\, \exp\left(-\frac{M_{Z_c(3900)}^2}{T^2}\right)+\lambda_{Z(4430)}^2M_{Z(4430)}^6\,
 \exp\left(-\frac{M_{Z(4430)}^2}{T^2}\right) \nonumber\\
 && =\int_{4m_c^2}^{s_0} ds\, s^3\rho(s) \, \exp\left(-\frac{s}{T^2}\right) \, .
\end{eqnarray}
We solve the coupled equations consistently and obtain the values of the masses of  the $Z_c(3900)$ and $Z(4430)$,
\begin{eqnarray}
M_{Z_c(3900)}&=&3.91^{+0.21}_{-0.17}\,\rm{GeV} \, ,  \,\,\, {\rm Experimental\,\, value} \,\,\,\,3899.0\pm 3.6\pm 4.9\,\rm{ MeV} \, \cite{BES3900}\,   , \nonumber\\
M_{Z(4430)}&=&4.51^{+0.17}_{-0.09}\,\rm{GeV} \, ,  \,\,\, {\rm Experimental\,\, value} \,\,\,\,4475\pm7\,{_{-25}^{+15}}\,\rm {MeV}\, \cite{LHCb-1404}\,   ,
\end{eqnarray}
which favors assigning  $Z_c(3900)$ and $Z(4430)$  to be  the ground state and the first radial excited state of the  $\bar{3}\otimes 3$ type axial-vector tetraquark states, respectively. For the technical details, one can consult Ref.\cite{Wang4430}. We can assign the $Z_c(4200)$ to be the  $8\otimes 8$ type axial-vector molecule-like state, or it is possible to assign the $Z_c(4200)$ to be the  $8\otimes 8$ type axial-vector molecule-like state.

In case II, the  $Z_c(4200)$ is  the ground state  of  the $\bar{3}\otimes3$ type  axial-vector  tetraquark state,
the $Z_c(3900)$ and $Z(4430)$ are not the $\bar{3}\otimes3$ type axial-vector tetraquark states, then the current couples potentially to the $Z_c(4200)$, $\langle 0|\eta_\mu(0)|Z_c(4200)\rangle=\lambda_{Z_c(4200)}\varepsilon_\mu$, where the $\varepsilon_\mu$ is  the polarization vector of the $Z_c(4200)$. Now we write down the QCD sum rule,
\begin{eqnarray}
\lambda_{Z_c(4200)}^2\, \exp\left(-\frac{M_{Z_c(4200)}^2}{T^2}\right) &=& \int_{4m_c^2}^{s_0} ds\, \rho(s) \, \exp\left(-\frac{s}{T^2}\right) \, .
\end{eqnarray}
 We differentiate   Eq.(48) with respect to  $\tau=\frac{1}{T^2}$,  then eliminate the
 pole residue  $\lambda_{Z_c(4200)}$ to obtain the QCD sum rules for the mass,
 \begin{eqnarray}
 M^2_{Z_c(4200)}= \frac{\int_{4m_c^2}^{s_0} ds\left(-\frac{d}{d \tau }\right)\rho(s)e^{-\tau s}}{\int_{4m_c^2}^{s_0} ds \rho(s)e^{-\tau s}}\, .
\end{eqnarray}
In Fig.7, we plot the mass of the $Z_c(4200)$  with variations of the  Borel parameters $T^2$ and energy scales $\mu$ for the threshold parameter $\sqrt{s_0}=4.7\,\rm{GeV}$. From the figure, we can see that the mass decreases monotonously with increase of the energy scales, the experimental value
$M_{Z_c(4200)} = 4196^{+31}_{-29}{}^{+17}_{-13} \,\rm{MeV}$ \cite{Zc4200exp} can be reproduced approximately  at the energy scales $\mu=(1.1-1.4)\,\rm{GeV}$. If the  $Z_c(4200)$ is  the ground state  of  the $\bar{3}\otimes3$ type  axial-vector  tetraquark state, we have to assign the $Z_c(3900)$ to be the $1\otimes1$ type molecular state \cite{WangHuang-molecule},
\begin{eqnarray}
Z_c(3900/3885)&=&\frac{1}{\sqrt{2}}\left( D\overline{D}^{*} +  D^{*}\overline{D}\right)\, ,
\end{eqnarray}
however, it is odd to assign the $Z(4430)$ to be  the excited $1\otimes1$ type molecular state.

\begin{figure}
 \centering
 \includegraphics[totalheight=8cm,width=12cm]{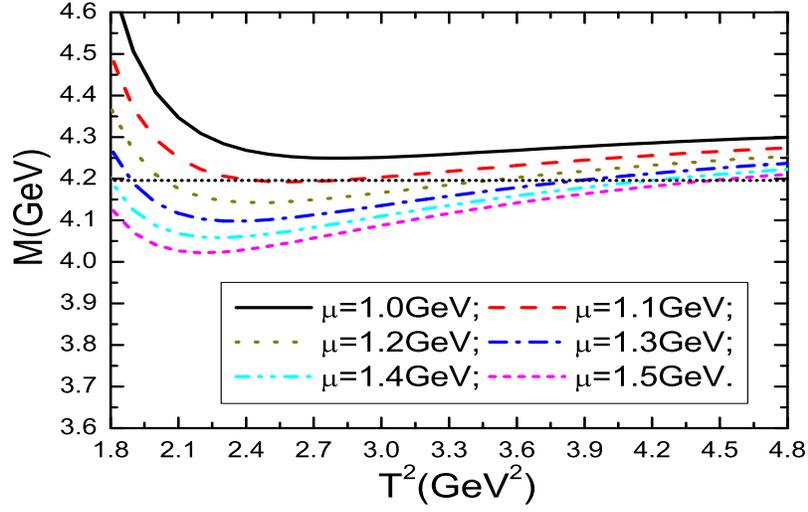}
    \caption{The mass  of the $Z_c(4200)$ with variations of the Borel parameters $T^2$ and energy scales $\mu$,  where  the horizontal line denotes  the experimental value.   }
\end{figure}

In case III, the  $Z(4430)$ is   the ground state  of the $\bar{3}\otimes3$ type  axial-vector  tetraquark state,
the $Z_c(3900)$ and $Z_c(4200)$ are not the $\bar{3}\otimes3$ type axial-vector tetraquark states, then the current couples potentially to the $Z(4430)$, $\langle 0|\eta_\mu(0)|Z(4430)\rangle=\lambda_{Z(4430)}\varepsilon_\mu$, where the $\varepsilon_\mu$ is  the polarization vector of the $Z(4430)$. Now we write down the QCD sum rule,
\begin{eqnarray}
\lambda_{Z(4430)}^2\, \exp\left(-\frac{M_{Z(4430)}^2}{T^2}\right) &=& \int_{4m_c^2}^{s_0} ds\, \rho(s) \, \exp\left(-\frac{s}{T^2}\right) \, .
\end{eqnarray}
 We differentiate   Eq.(51) with respect to  $\tau=\frac{1}{T^2}$,  eliminate the
 pole residue  $\lambda_{Z(4430)}$ to obtain the QCD sum rules for the mass,
 \begin{eqnarray}
 M^2_{Z(4430)}= \frac{\int_{4m_c^2}^{s_0} ds\left(-\frac{d}{d \tau }\right)\rho(s)e^{-\tau s}}{\int_{4m_c^2}^{s_0} ds \rho(s)e^{-\tau s}}\, .
\end{eqnarray}
In Fig.8, we plot the mass of the $Z(4430)$  with variations of the  Borel parameters $T^2$ and energy scales $\mu$ for the threshold parameter $\sqrt{s_0}=5.0\,\rm{GeV}$. From the figure, we can see that the experimental value $M_{Z(4430)}= 4475\pm7\,{_{-25}^{+15}}\,\rm {MeV}$ \cite{LHCb-1404} can be reproduced approximately at the energy scale $\mu=1\,\rm{GeV}$ \cite{Wang4430-1GeV}. At the energy scales $\mu > 1\,\rm{GeV}$, the predicted mass $M_{Z(4430)}$ is much smaller than $4475\,\rm{MeV}$, assigning  the $Z(4430)$ to be  the ground state  of the $\bar{3}\otimes3$ type  axial-vector  tetraquark state is not favored.

\begin{figure}
 \centering
 \includegraphics[totalheight=8cm,width=12cm]{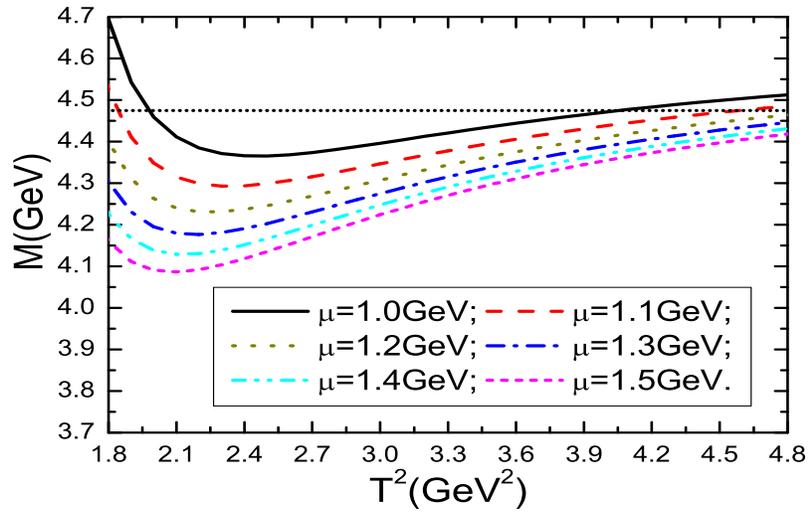}
    \caption{The mass  of the $Z(4430)$ with variations of the Borel parameters $T^2$ and energy scales $\mu$,  where  the horizontal line denotes  the experimental value.  }
\end{figure}

\section{Conclusion}
In this article, we assume the $Z_c(4200)$ as the $8\otimes8$ type axial-vector molecule-like state, and construct the  $8\otimes8$ type axial-vector
current to study its mass and width with the QCD sum rules.   The numerical result  supports assigning  the   $Z_c(4200)$  to be  the $8\otimes8$  type  molecule-like  state with $J^{PC}=1^{+-}$. Furthermore, we discuss the possible assignments of the $Z_c(3900)$, $Z_c(4200)$ and $Z(4430)$ to be the $\bar{3}\otimes 3$ type tetraquark states with $J^{PC}=1^{+-}$.

\section*{Acknowledgements}
This  work is supported by National Natural Science Foundation,
Grant Number 11375063, and Natural Science Foundation of Hebei province, Grant Number A2014502017.

\end{document}